\documentclass[reprint,superscriptaddress,noeprint,nolongbibliography]{revtex4-2}

\usepackage{graphicx}
\usepackage{float}
\usepackage{dcolumn}
\usepackage{bm}
\usepackage{tikz}
\usepackage{enumitem}
\usepackage{amsfonts}
\usepackage{amsmath}
\usepackage{xcolor}
\usepackage{soul}
\usepackage[export]{adjustbox}

\usepackage[normalem]{ulem}

\expandafter\def\expandafter\normalsize\expandafter{%
    \normalsize%
    \setlength\abovedisplayskip{0pt}%
    \setlength\belowdisplayskip{8pt}%
    \setlength\abovedisplayshortskip{-8pt}%
    \setlength\belowdisplayshortskip{2pt}%
}
\DeclareMathOperator\erfc{erfc}
\newcommand{\olsi}[1]{\,\overline{\!{#1}}}

\begin{document}


\title{Stacked Rayleigh-Taylor instabilities grow drops into soft stalactite-like structures}



\author{Barath Venkateswaran}
\affiliation{Department of Chemical and Biological Engineering, Princeton University, NJ 08540, USA}
\author{Trevor J. Jones}
\affiliation{Department of Mechanical Engineering, Carnegie Mellon University, Pittsburgh, PA 15213, USA}
\author{Grace Kresge} 
\affiliation{Department of Chemical Engineering and Materials Science, Minneapolis, MN 55455-0132, USA}
\author{Joel Marthelot}
\affiliation{CNRS, IUSTI, Aix-Marseille Universite, 13013 Marseille, France}
\author{Etienne Jambon-Puillet}
\affiliation{LadHyX, CNRS, Ecole Polytechnique, Institut Polytechnique de Paris, Palaiseau, France}
\author{P.-T. Brun}
\affiliation{Department of Chemical and Biological Engineering, Princeton University, NJ 08540, USA}


\date{\today}

\begin{abstract}

The interplay between thin film hydrodynamics and solidification produces formidably intricate geophysical structures, such as stalactites and icicles, whose shape is a testimony of their long growth. In simpler settings, liquid films can also produce regular patterns. When coated on the underside of a flat plate, these films are unstable and yield lattices of drops following the Rayleigh-Taylor instability. While this interfacial instability is well-studied in Newtonian fluids, much less is known about what happens when the thin film solidifies. 
Here, we coat the underside of a surface with liquid elastomer, allowing the film to destabilize and flow while it cures into an elastic solid. Once the first coating yields an array of solid droplets, this iterative coat-flow-cure process is repeated and gives rise to corrugated slender structures, which we name flexicles for their resemblance to icicles. We study the subtle combination of chaos and order that confers our flexicles, their structure, shape, arrangement, and, ultimately, deformability. 

\end{abstract}
\maketitle

In nature, growth and pattern formation go hand-in-hand to produce intricate morphologies. While prominent in living organisms~\cite{thompson_1992}, complex patterns are also ubiquitous in inorganic settings. Examples include icicles, stalactites, and draperies whose geomorphic structures develop as matter is continuously added via freezing or deposition~\cite{short2005, Ueno2007, Chen_2013,ladan_morris_2022}. Solidification is typically coupled to the flows in the liquid films that coat such structures, which are known to be unstable when on the underside of a substrate~\cite{stoneRTI2014,gallaire_brun_2017,balestra_kofman_brun_scheid_gallaire_2018}. 
The Rayleigh-Taylor instability (RTI) in a film is a classic fluid dynamics problem where the tug-and-pull between the stabilizing interfacial tension and destabilizing gravity produces droplets in coatings~\cite{fermigier1992,de2004capillarity}. These droplets, whose characteristic length is well captured by linear stability analysis~\cite{chandrasekhar1961,drazin2002introduction,gallaire_brun_2017}, are arranged in various patterns of different symmetries~\cite{fermigier1992}. Although the details of the nonlinear selection mechanism remain elusive, these two-dimensional lattices can be controlled using templates and seeds~\cite{fermigier1992,joel2018}. 
As time progresses, dripping typically occurs~\cite{lister2010}, progressively diminishing the number of suspended droplets. Curable liquids interrupt dripping and form corrugated solids whose shape is imparted by the interplay between flow and solidification~\cite{brun2022}. While well understood, the amplitude of these artificial solids is far smaller than that of their geophysical counterparts, which develop over longer periods of time~\cite{short2005,Ueno2007,Chen_2013,ladan_morris_2022}. 

\begin{figure}[!h]
    \centering
\includegraphics[width=.5\textwidth]{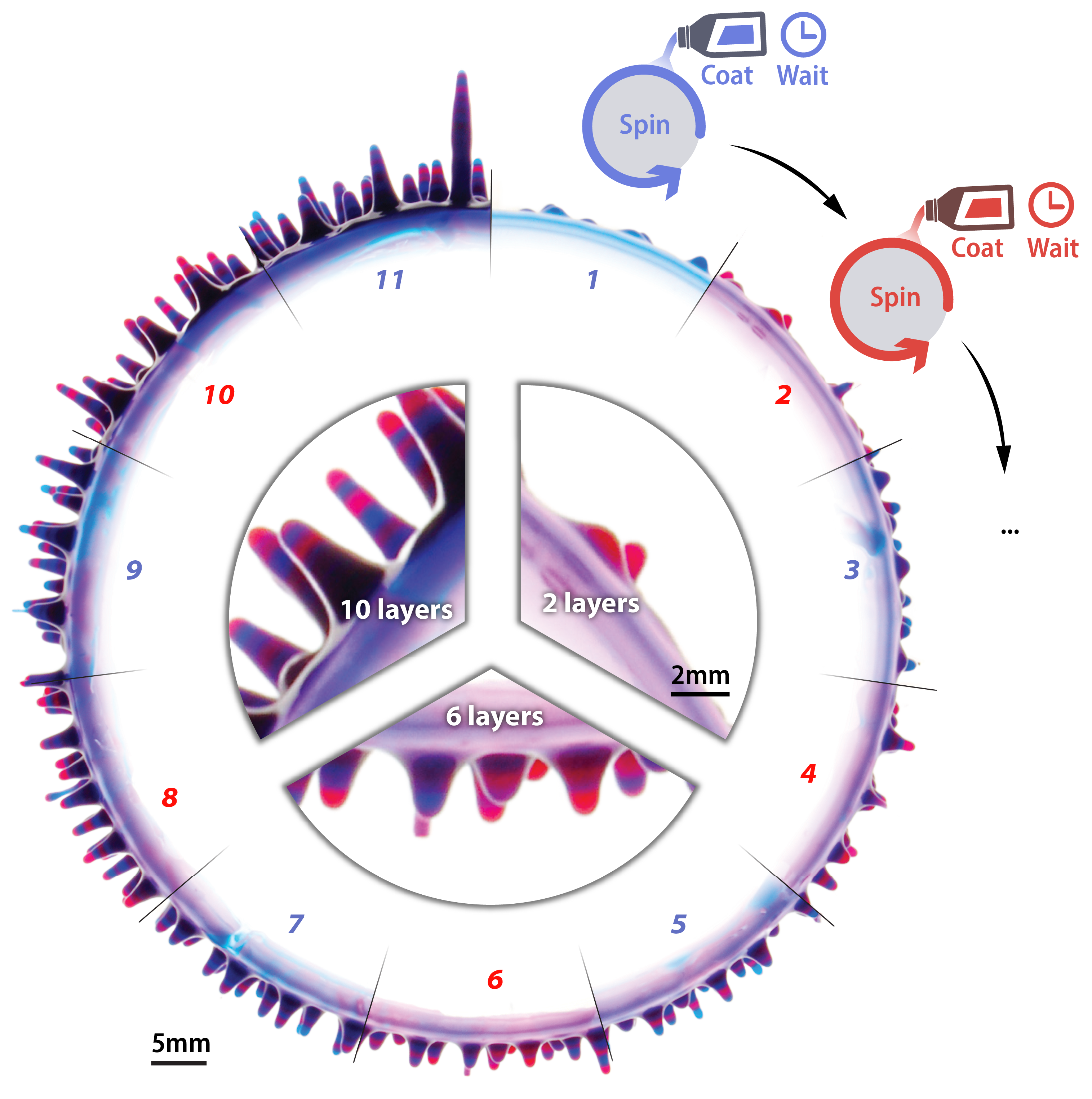}
    \caption{\textbf{Iterative growth of flexicles}. Stitched photographs of a cylinder with radius $R=3.75$ cm sequentially coated with curable elastomer while being rotated with speed $\Omega = 350$ rpm ($R\Omega^2 = 5.1$g). The number of coatings ranges from 1 to 11 and increases clockwise. 
    }
    \label{fig:fig1}
\end{figure}

In this Letter, we study the progressive development of patterns of stalactite-like elastomeric structures, which we call flexicles. As reported in Fig.~\ref{fig:fig1}, we apply a series of coatings of curable polymers to a substrate. For each coating, the fluid polymer flows by the action of an acceleration field and is allowed to cure before the subsequent coating is applied. The coatings are applied to the outside of a rotating cylinder or the underside of a flat plate so that the acceleration fields are centrifugal and gravity, respectively. In any case, we operate in a regime where the instability develops much faster than the time $\tau_c$ needed for the polymer to cure. In fact, the time scale of the instability is $12\mu_0\ell_c^4/\gamma h_0^3\ll \tau_c$, with $\mu_0$ the initial viscosity of the polymer, $\gamma$ its surface tension, $h_0$ the typical thickness of a coating and $\ell_c=\sqrt{\gamma/\rho a}$ the capillary length of the problem, with $\rho$ the liquid density and $a$ the magnitude of the acceleration field~\cite{joel2018,jambon2021elastic}. The first coating produces drops distributed over the substrate, which are sufficiently small to stay on the surface as they cure~\cite{joel2018}. As this coat-flow-cure process is repeated, the solid surface formed from previous coatings acts as the substrate for the subsequent coating, thereby allowing the overall structure to grow, as evident from the inset of Fig.~\ref{fig:fig1}. 
This process produces slender, tapered structures whose growth and properties are rationalized in this Letter.


\begin{figure}[t]
    \centering
    \includegraphics[width=0.5\textwidth]{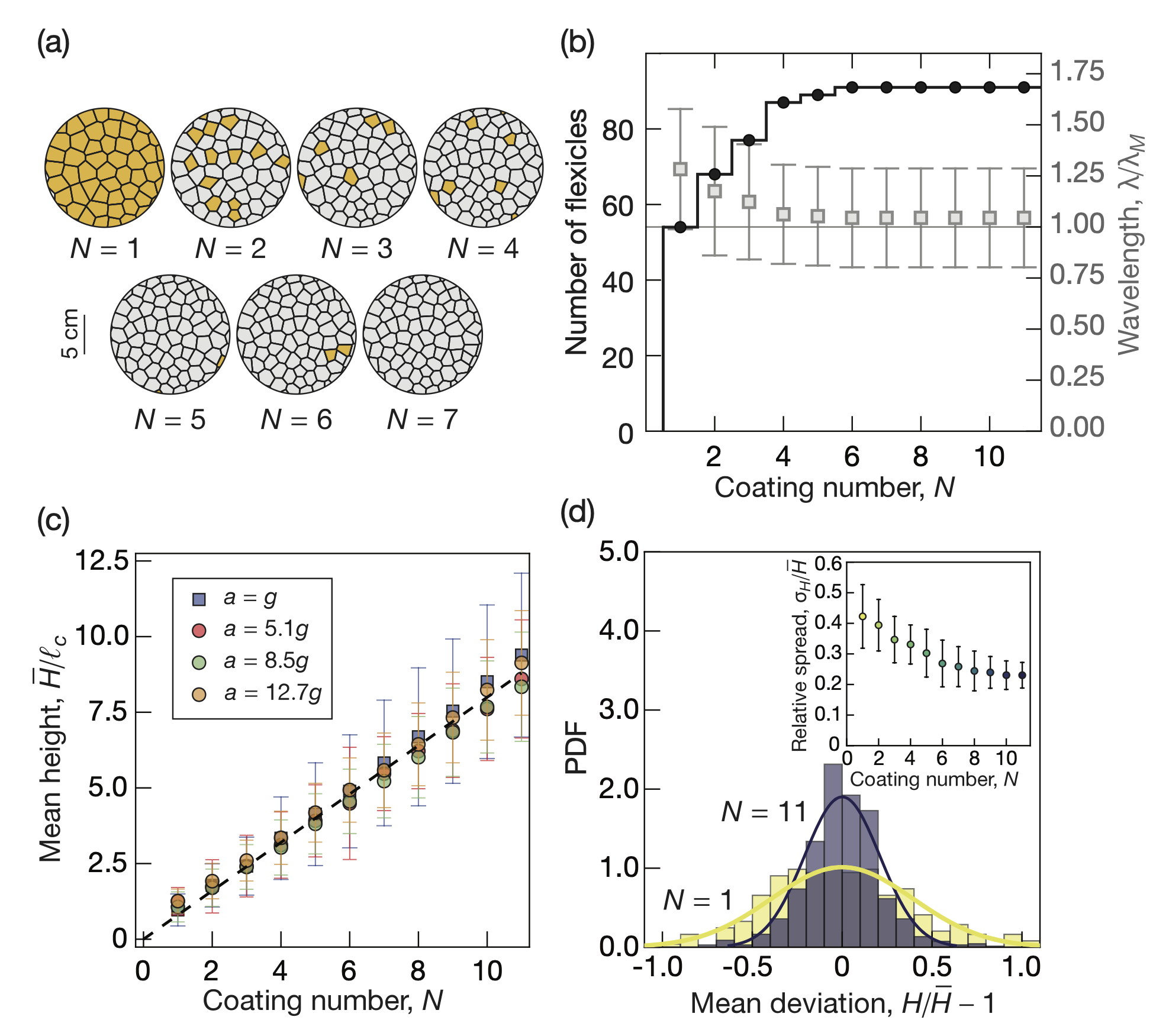}
    \caption{\textbf{Pattern growth}. (a) Voronoi tessellation of the solid drops' peaks formed under gravity after each coating. Yellow tiles represent the newly formed drops and grey tiles represent the previously formed drops. (b) Number of flexicles and wavelength as a function of the coating number $N$. (c) Height of the flexicles as a function of $N$ (dashed line is a linear fit denoting the average height). The squares represent structures formed by gravity and the circles denote the structures formed by centrifugation. (d) Probability density function of the normalized mean deviation of flexicle peak height, $H/\overline{H} - 1$, for the first and last coating ($N=1\text{ and } 11$), using data from all the acceleration fields mentioned in (c). Inset shows the relative spread of the flexicle peak heights for different number of coatings.
    }
    \label{fig:fig2}
\end{figure}

In Fig.~\ref{fig:fig2}a, we show the Voronoi tessellations that describe the position of our structures' peaks in the case where coatings are applied on a flat surface (see Methods~\ref{meth:fabric}; $a=g$ and $\ell_c = \sqrt{\gamma/\rho g}$). The tiles in yellow indicate where new structures appear at each coating.  While the first instance of the RTI yields the greatest number of new drops, we find a sizable number of drops still appear in subsequent coatings. These new drops occupy the space between existing structures, rearranging the whole pattern tessellation (See SI). In Fig~\ref{fig:fig2}b, we report this data and show the number of flexicles progressively increases and plateaus after six coatings. At this point, we find that the distance between flexicles matches $\lambda_M=2\pi\sqrt{2}\ell_c$, the most unstable mode of the RTI, such that the results of linear stability remain relevant to the problem even far from the instability threshold. However, our tesselations differ from that of a perfect hexagonal arrangement, usually thought to be the most unstable pattern~\cite{fermigier1992}. In contrast, we report an average shape factor~\cite{reis2015}, $\xi =1.194\pm0.06$, where $\xi=p_i^2/(4\pi A_i)$, with $p_i$ and $A_i$ being the perimeter and surface area of each tile. Our tiles are thus between a regular pentagon ($\xi_5=1.156$) and a square ($\xi_4=1.273$). This spatial arrangement is preserved for further coatings while the flexicles grow.  

In Fig.~\ref{fig:fig2}c, we report the average height of our structures, $\olsi{H}$, versus the number of coatings $N$ for four different acceleration fields ($a=g$ and $a=R\Omega^2$ with $R$ and $\Omega$ the substrate radius and rotation speed). Altering the acceleration fields allows us to produce structures with different length scales. After eleven coatings, gravity alone yields structures with mean height $\olsi{H}\simeq\,$1cm. Conversely, we have flexicles an order of magnitude smaller, $\olsi{H}\simeq$mm, for rotating samples with acceleration $a=12.7 g$.  Yet, these data collapse onto a single master curve when rescaled with the capillary length, $\ell_c$, calculated for each experiment. The mean height grows as:

\begin{equation}
\label{eq:1}
    \olsi{H} = (0.8 \pm 0.02) N\ell_c
\end{equation}

between coatings. This linear relation further confirms the relevance of the tug-and-pull between interfacial forces and the destabilizing acceleration field used to define $\ell_c$. As evident from error bars in Fig.~\ref{fig:fig2}c, the raw variability of $H$ seems to increases with $N$. We now inspect the height distribution of flexicles found in our samples more closely.

In Fig.~\ref{fig:fig2}d, we present two histograms showing the probability density function of the normalized mean deviation of the structures' height, $H/\olsi{H}-1$. The distributions become narrower as the number of coatings increases, indicating that every new layer reduces the height variability relative to $\olsi H$. The variability seen in the initial few coatings does not have a persisting effect on the structures at later coatings. Instead, the relative spread decreases and appears to saturate to $0.2$ (see inset). Some regularity thus emerges from the apparent turmoil of the experiment: excess volumes of fluid are applied, yielding to rapid evolution and dripping until the coatings are thin enough and stabilize to the observed flexicles. 
So far, we discussed the pattern properties and the structure's heights. To rationalize their shape, we now examine our flexicles' internal cross-sections aiming to elucidate the physics at play in the pattern's growth.
 
\begin{figure*}[t]
    \includegraphics[width=.85\textwidth]{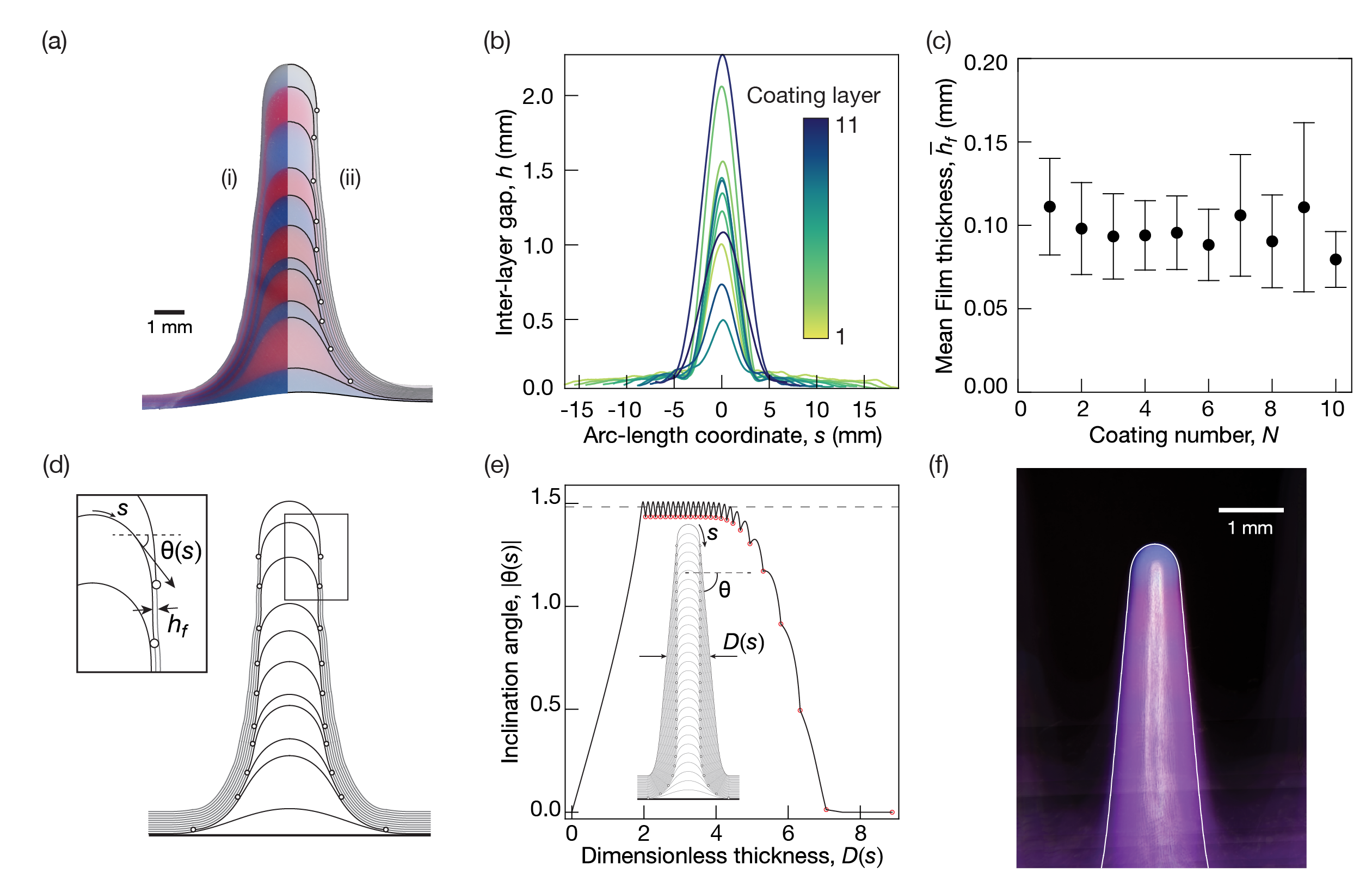}
        \caption{\textbf{Internal structure of a flexicle.} (a) (i) Sample cross-section with (ii) thin-film and pendant drop regions highlighted using image analysis. Black circles indicate the transition point between the two regions in each layer ($a=g$). (b) Inter-layer gap as a function of $s$ for 11 coatings ($a=g$) for a single flexicles. (c) Mean film thickness in each layer, averaged across the sample. (d) Reconstructed sample using our model (see Eq.~\eqref{eq:pendantode} and \eqref{eqn:layer_1} in Methods). (e) Theoretical prediction of the inclination of the surface, $|\theta(s)|$ as a function of the thickness after 25 coatings, $D(s)$ (expressed in $\ell_c$ units). The dashed line represents the cone solution. Red circles mark the locus of corrugations. (f) RGB-averaged image of 32 flexicles shifted so that their apexes match ($R\Omega^2 = 12.7g$). The white curve is the solution in (e).}
    \label{fig:fig3}
\end{figure*}


In Fig~\ref{fig:fig3}a(i), we show a photograph of the cross-section of a typical flexicle formed under gravity. The alternating red and blue colors represent successive coatings. We find that each layer comprises a thin film (shown in gray in Fig~\ref{fig:fig3}a(ii)), which connects to a thicker pendant drop region (shown in black). This transition occurs at a distance $s\simeq \pm 5$ mm from the apex as evident in Fig.~\ref{fig:fig3}b, where we report the inter-layer gap, $h$, defined as the Euclidian distance to the previous layer. We also observe no apparent correlation between the coating number and the magnitude of $h$ at the apex. Large and small drops alternate, depending on the volume of the residue left after the excess elastomer drips.  In contrast, the film thickness decreases continuously by drainage~\cite{jones2021}, yielding a remarkably uniform layer across all the coatings. In Fig.~\ref{fig:fig3}c, we report the film thickness averaged across all the flexicles in a given experimental sample and find that $\bar{h}_f \approx 0.1$ mm ($\star$). 

Upon observing the nature of the cross-sections of our structures, we propose a simple model to capture their growth. We hypothesize that the drop regions of our samples have reached a steady-state shape that is solidified as curing occurs. As such, we find the shape of the first layer by solving the axisymmetric Young-Laplace equation for a pendant drop (see Methods~\ref{meth:mmtk}). To build any subsequent layer $N$, we proceed in two steps. We first shift the previous layer $N-1$ normal to the surface by a distance set to the mean film thickness. We then find the shape of the drop region by solving Young-Laplace's equation, using boundary conditions that match the thin-film region smoothly. This problem has a continuous range of solutions with different heights. In Fig.\ref{fig:fig3}d, we show the results obtained with our model when choosing the successive drops' height so that they match experimental values in Fig.\ref{fig:fig3}a. The shapes we obtain by integrating our model are close to those seen in the experiment. 

In search of universality, we now build an average flexicle by inputting into our model a unique drop height and film thicknesses, which are consistent with eqs.~(\ref{eq:1}), and ($\star$). The results are shown in Fig~\ref{fig:fig3}e, where we plot the flexicle slope as a function of its dimensionless thickness $D(s)$. We find that the slope is uniform in the central region of the flexible, forming a cone with an opening angle approximatively equal to $5 ^\circ$ (dashed line in Fig.\ref{fig:fig3}e). Note the small amplitude oscillations on the surface that echo the ripples typically seen on icicles~\cite{Chen_2013,ladan_morris_2022}. To test this model, we compare our prediction with a composite image of a collection of tip-aligned flexicles (in Fig.~\ref{fig:fig3}f, the image shows the mean RGB values). We find a favorable agreement between the two, allowing us to conclude that flexicles are truncated cones topped off with a pendant drop. 

\begin{figure}[t!]
    \includegraphics[width=0.5\textwidth]{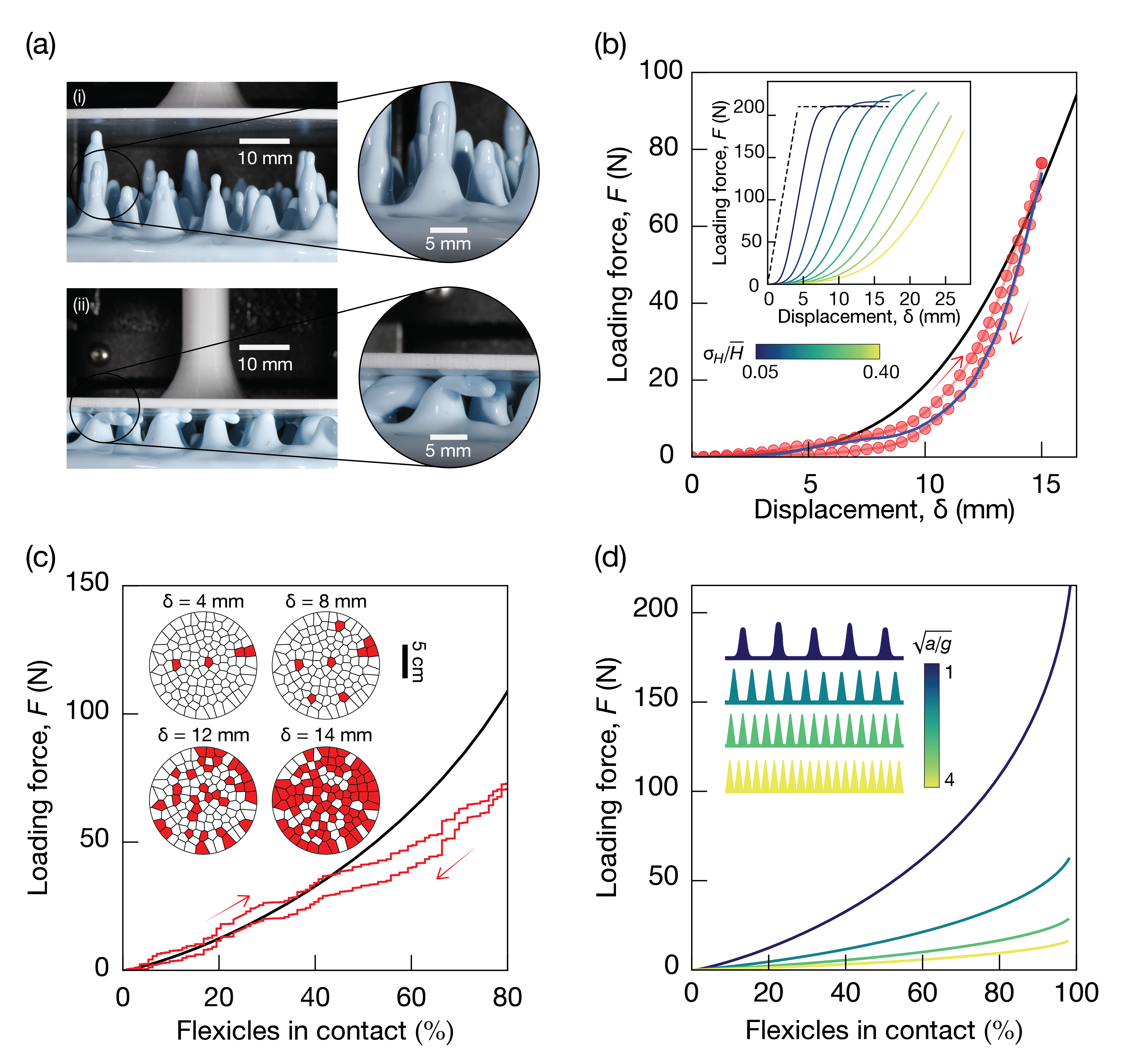}
    \caption{\textbf{Indentation response of a collection of flexicles.} (a) Snapshots of the indentation of 113 flexicles  (15 layers). (b) Loading force data (red) vs indenter displacement. Predictions are shown in blue (exact) and black (statistical). Inset shows the variation of the latter when varying the dispersity of flexicles' heights (loading is interrupted when reaching $80\%$ of the expected maximum height). (c) Loading force vs fraction of flexicles in contact with the indenter (data in red and statistical model in black). Inset shows the flexicles in contact as indentation increases. (d) Loading force profiles for flexicles of the same mean height and produced with different accelerations (see inset for a sketch).
}
    \label{fig:fig4}
\end{figure}

In this last section, we examine the deformability of such structures. To this end, we indent an array of flexicles (see Methods). Fig.~\ref{fig:fig4}a(i) shows the sample when the indenter first makes contact with the tallest flexicle while others remain free. Fig.~\ref{fig:fig4}a(ii) shows the sample when most flexicles are deformed, typically bending in the same direction, a self-ordering promoted by interactions in this confined configuration~\cite{guerra2023self}. 
In Fig.~\ref{fig:fig4}b, we show the corresponding loading force $F$ as a function of the imposed displacement of the indenter $\delta$. The response amounts to a concave-up curve, which reflects that more and more flexicles are deformed as indentation proceeds (see Fig.~\ref{fig:fig4}c), thus increasing the apparent stiffness of the sample. Individual flexicles are modeled as compressible clamped-hinged Kirchhoff rods that show a linear elastic response, which then saturates to a force plateau when buckling occurs (see Methods). The surface response is the sum of these individual piecewise functions. When feeding the model with the measured height of each flexicle in the sample, our prediction for $F(\delta)$ (blue curve in Fig.~\ref{fig:fig4}b) compares favorably with experiments (red points), indicating the validity of our reduced-order model and the approximations made to account for the deformability of individual flexicles. The black curve indicates the force response from a statistical version of the model, which uses a Gaussian distribution for the flexicles' heights. The distribution uses the previously obtained mean height, $0.8N\ell_c$, with a 30\% standard deviation ($N = 15$ and $\ell_c = 1.41$ mm, see Methods). The agreement between theory and experiments is fair. Note that the sample's dispersity plays a key role in setting the stiffness of the flexicles' ensemble. Monodisperse flexicles lead to a ramp-like response (see inset of Fig.~\ref{fig:fig4}b), while increasing the dispersity softens the elastic response.  

Fig.~\ref{fig:fig4}c shows how such flexicles could be leveraged as rudimentary force sensors, where the force is estimated by enumerating the flexicles in contact with a given substrate. The figure shows that the force recorded during indentation varies almost linearly with the fraction of flexicles in contact with the indenter in the range explored. Forces of about $50$ N can be accurately estimated ($\pm 0.5 $N) thanks to the large number of discrete steps observed as force increases (see red line). The value of the maximum attainable force can be lowered, and thus, the method's sensitivity increased by employing thinner flexicles obtained with larger acceleration fields. Fig.~\ref{fig:fig4}d shows how the elastic response can be adjusted by varying the aspect ratio of the flexicles. Thus, our methodology and models allow us to tune the stiffness of a collection of thin elastic structures that possess a controlled degree of polydispersity.

In this Letter, we leveraged the Rayleigh-Taylor instability in curable elastomers to produce soft, slender structures by stacking instances of the thin-film instability. We demonstrated through experiments that, although the formation of these flexicles seems disordered, the average height of the structures grows linearly with the number of coatings converging to a conical shape capped by a pendant drop; and that although we are far from threshold, the arrangement of these structures on the surface is well captured by linear stability analysis arguments. While the growth of our flexicles tends to converge (the relative deviation to the mean decreases), our samples still show appreciable heterogeneities, which have direct consequences on the surface properties, e.g., on its elastic properties. 

\section*{Acknowledgments}
We thank H. Stone for comments and feedback. This work was supported partially by the National Science Foundation under Grants no. CBET 2042930 and Grants no. CMMI 2037097.


%

\newpage
\section*{Methods}
\subsection{Flexicle fabrication}
\label{meth:fabric}
We use polyvinylsiloxane (VPS; Zhermack Elite Double 8) to produce our structures. To initiate the curing process, the base and curing agent are mixed in a 1:1 weight ratio using a centrifugal mixer (Thinky ARE-310) for 10s clockwise (2000 rpm) and 10s anti-clockwise (2200 rpm) at room temperature (20$^{\circ}$C). These mixtures tend to display a snap-set behavior where they exhibit nearly constant viscosity for several minutes after mixing followed by rapid crosslinking to form elastic materials. The time-dependence of the elastomers’ viscosity is well described by the equation $\mu = \mu_0(1-t/\tau_c)^{-n}$. The key parameters are the initial/working viscosity $\mu_0$ and a characteristic curing time $\tau_c$. As for the exponent, $n = 2$ fits fairly well with viscosity measurements of various VPS mixtures. $\mu_0$ is typically around 1.5 Pa.s and $\tau_c$ is around 10 min~\cite{jones2021}.

For experiments where gravity is the destabilizing body force, we take a flat, horizontal plate made of acrylic (Radius =  and apply a coating of VPS-08. The first coating alone is made using a spin coater (Laurell WS-650). We then flip the plate to initiate the RT instability and dripping to occur. Once the elastomer coating cures, we obtain a lattice of elastic drops. We pour another coating of VPS-08 with excess volume on the cured surface and allow it to drip and cure as before. This process of coat-flow-cure is repeated with alternating silicone dyes (Silc Pig) for visualization purposes.

To mimic higher gravity conditions, we use centrifugal force as our destabilizing body force and repeat the above process with a few modifications. Instead of coating and flipping a flat plate, we coat the curved surface of a cylinder made of acrylic and spin it about its axis using a spin coater (Laurell WS-650). The radius of the cylinder, $R = 3.75$ cm is chosen such that the effect of cylinder curvature is less significant than the capillary effects ($R>>\ell_c$). The new length scale of the system is now tunable with the rotation speed, $\Omega$ as $\ell_c =\ell_{c,g} \sqrt{g/R\Omega^2} $, where $\ell_{c,g} = \sqrt{\gamma/\rho g}\approx 1.414\text{ mm}$ is the length scale of the experimental system when gravity destabilizes a coating of VPS-08. We use different rotation speeds $\Omega = \{350,450,550\}$ rpm to alter the length scales of structures.

\subsection{Modeling flexicle shapes using \textit{Mathematica}}
\label{meth:mmtk}
We model this highly non-linear process of stacking instabilities on top of each other by making a number of assumptions. We assume that every layer is made up of a uniform, thin film, and a pendant drop. We sequentially build a composite layer of thin film and pendant drop on top of previously built layers. To build a new layer, we match the hydrostatic, pendant drop solution to a uniform thin film solution along the surface of the existing solution. We then compare our model solutions with the actual structures formed by a combination of hydrodynamics and solidification.

Using a hydrostatic pressure balance, we obtain an ODE \eqref{eq:pendantode} which describes all possible equilibrium drop shapes. We define $L$ as the half length of the drop and use it to rescale the arc-length $s$ such that it varies between $0$ and $1$. The drop surface cylindrical coordinates $(r,z)$  are rescaled with $\ell_c$. Assuming total wetting since the drop connects to a thin film (zero contact angle), yields Eq.(1) that we integrate numerically with Mathematica.
\begin{gather}
     \theta''(s) = -\tilde{L}^2\sin\theta(s) - \tilde{L}\left(\frac{\sin\theta(s)}{r(s)}\right)'\nonumber\\
    r'(s)=\tilde{L}\cos\theta(s),\ z'(s) = \tilde{L}\sin\theta(s)\quad \forall\  s\in[0,1] 
    \label{eq:pendantode}
\end{gather}

where $\theta(s)$ is the angle made with the horizontal by the tangent to the curve and $\tilde{L}=L/\ell_c$.

Along with the second-order pendant drop ODE in equation \eqref{eq:pendantode}, we have two first-order ODEs for the parametric representation of drop shape and one unknown, $\tilde{L}$ which requires 5 BCs in total. For the first coating, we use the usual boundary conditions for a pendant drop on a flat plate:
\begin{gather}
       \theta(0) = 0,\ z(0) = 0,\nonumber\\
    \theta(1) = 0,\ r(1) = 0,\ z(1) = A_1 
    \label{eqn:layer_1}
\end{gather}
where $A_1$ is the height of the drop.  Equations \eqref{eq:pendantode} and \eqref{eqn:layer_1} together constitute a boundary-value problem. Previous works like~\cite{pitts_1973,pitts_1974,de2004capillarity,sumesh2010,joel2018} contain discussions on possible pendant drop shapes. 


From here, we attempt to stack another pendant drop solution on the previously formed curved solution (henceforth called 'base layer'). We model the new layer as a combination of a pendant drop and a uniform thin film; the pendant drop starts where the thin film ends. To ensure a smooth transition between regions, we look for solutions of  \eqref{eq:pendantode} where the beginning of the curve ($s=0$) is parallel to the base layer, and is spaced a distance $h_f$ normally from it, where $h_f$ is the prescribed thickness of the thin-film region ($h_f\approx0.07\ell_c$ from experiments). 
Define $s^*\in(0,1)$ as the arc-length coordinate of the base layer that corresponds to the point on the base layer from which the new solution is anchored (with a normal offset of $h_f$). We solve for the new solution by employing a shooting method from $s=0$ to $s=1$. We use $\{s^*, \theta'\}$ as shooting parameters to get a solution such that at $s=1$, we have $r = 0, \theta = 0 \text{, and } z = A_2$. From the second layer onwards, we have two unknowns to solve for $\{s^*, \tilde{L}\}$ along with the same second order pendant drop ODE and two first order ODEs as in equation \eqref{eq:pendantode}.  So, we now need 6 BCs, which are taken care of by the $\{r,z,\theta\}$ values at either end. We repeat this procedure for as many layers as we require, by reassigning the penultimate layer as the base layer at the start of each iteration.

\subsection{Force measurements using \textit{Instron}}
\label{meth:instron}
We subject the carpet of flexicles produced after coat-flow-cure iterations with VPS-08 elastomer to an indentation test. We use flat indenters attached to a load cell for the tests. The indenters will press down against the fixed sample and record the loading forces as a function of deformation. We use the Instron 5944 - Single Column Tabletop Universal Testing System along with the Instron 2580 Series - 2kN Static load cell to record loading forces. Two kinds of indentation tests are performed: (1) collective deformation, using a circular indenter as wide as the flexicle surface, and (2) individual deformation, using a circular indenter narrow enough to deform one flexicle at a time.

\subsection{Modeling flexicles' response to indentation}
\label{meth:force}


Traditional Kirchhoff rod equations are valid for modeling the deformation of thin and inextensible rods. Define $S$ as the arc-length coordinate along the undeformed rod of length $H$ and radius $R$. For a rod with centerline described by $\mathbf{r}(S)$, we introduce rod internal tension $\mathbf{n}$, internal moment $\mathbf{m}$, and body forces $\mathbf{f}$. Assuming the rod lies on the $x$-$y$ plane, $\mathbf{n} = (n_x,n_y,0)$ and $\mathbf{m}=(0,0,m)$. The Kirchhoff rod equation then reads:

\begin{subequations}

\begin{equation}
    \frac{\partial\mathbf{n}}{\partial S} + \mathbf{f} = 0
\end{equation}

\begin{equation}
    \frac{\partial\mathbf{m}}{\partial S} + \frac{\partial \mathbf{r}}{\partial S}\times\mathbf{n} = 0
\end{equation}
\label{SIeq:kirchhoff}
\end{subequations}

We now allow for small compressions along the length of the rod. By taking $R \ll H$, we assume that any change in radius is negligible compared to the change in length. Let $s$ be the arc-length coordinate of the deformed rod. 
The longitudinal tension $(n_x \cos\theta  + n_y\sin\theta)$ is related to the length-change factor, $\lambda =  \frac{\partial s}{\partial S}$ as, 

\begin{equation}
   (n_x \cos\theta  + n_y\sin\theta) = EA(\lambda - 1)
\label{SIeq:stretcheq}
\end{equation}

where $E$ is the Young's modulus and $A = \pi R^2$ is the rod's cross sectional area. 

The moment $m$ has the following constitutive relation:

\begin{equation}
    m = EI\lambda\frac{\partial \theta}{\partial s}=EI\frac{\partial \theta}{\partial S}
\label{SIeq:momenteq}
\end{equation}

where $I = \pi R^4/4$ is the rod's moment of inertia. 

In our indentation experiments, $\mathbf{f}=0$. So, $n_x$ and $n_y$ are constants. Using the characteristic quantities, $x,y,S \sim H$, $m \sim EI/H$ and $n_x,n_y \sim EI/H^2$, we non-dimensionalise above Eqs.~\eqref{SIeq:kirchhoff},\eqref{SIeq:stretcheq} and \eqref{SIeq:momenteq} to get a set of first-order ODEs for $S\in[0,1]$:

\begin{align}
    \theta' &= m \nonumber\\
    m' &=\lambda\left(n_y\cos\theta - n_x\sin\theta\right)\nonumber\\
    x' &=\lambda\cos\theta\nonumber\\
    y' &=  \lambda\sin\theta\nonumber\\
    n_x'&=0\nonumber\\
    n_y'&=0
\label{SIeq:comprod}
\end{align}

where $\lambda=1 + \left(n_x \cos\theta  + n_y\sin\theta\right)/\beta$ and $\beta= \frac{EA}{EI/H^2} = 4 H^2/R^2$ after non-dimensionalisation. Denote $(.)' = \frac{\partial (.)}{\partial S}$. $|n_y|$ is a constant and is the vertical loading force on the rod. Note that, for $\beta\to\infty$, the above equations reduce to the traditional Kirchhoff equations for inextensible rods.

For the classic case of a clamped-hinged incompressible rod, the first Euler buckling mode corresponds to a loading force of $F_e=\zeta_1^2 \frac{EI}{H^2}$, where $\zeta_1\approx 4.493$ is the first non-zero root of $\tan \zeta = \zeta$. We approximate the slender flexicle as a soft, compressible rod of length $H$ and radius $R$ that is clamped at the bottom and that is hinged at the top. The boundary conditions for the case where an indenter displaces the hinged-top of the rod by $\delta$ are as follows:

\begin{align}
    x(0) = 0,\ y(0) = 0,\ \theta(0) &= \pi/2\nonumber\\
    m(1) = 0,\ x(1) = 0,\ y(1) &= 1-\delta/H
    \label{SIeq:comprodbc}
\end{align}

\begin{figure}[t]
    \centering
    \includegraphics[width=0.5\textwidth]{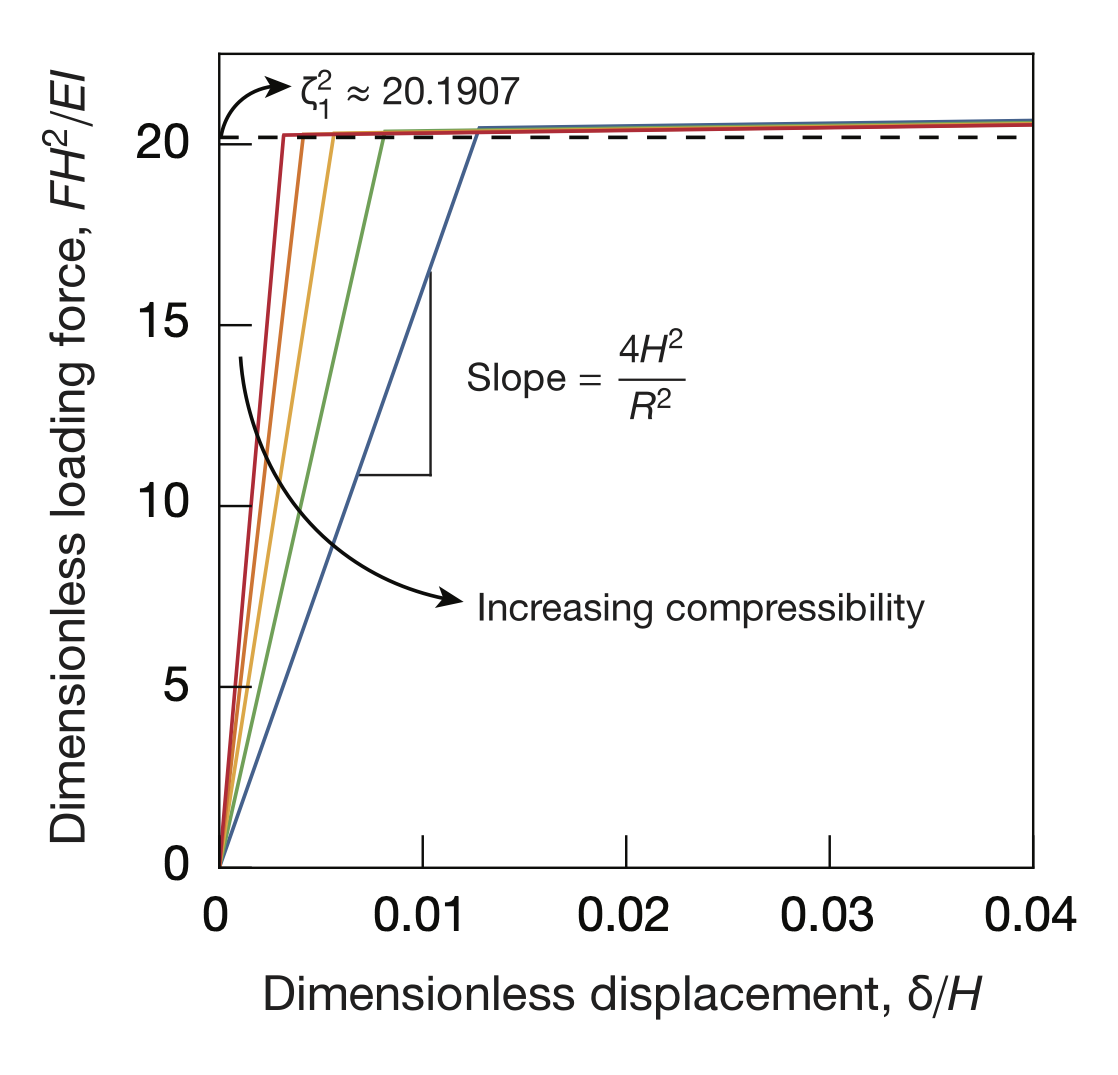}
    \caption{Typical numerical solutions of \eqref{SIeq:comprod} an \eqref{SIeq:comprodbc}. The dashed line indicates the Euler buckling load corresponding to an incompressible clamped-hinged rod - $\zeta_1$ is the first non-zero root to $\tan\zeta = \zeta$}
    \label{SIfig:numericalsoln}
\end{figure}

Using \textit{Mathematica}, we numerically solve the above set of equations for different values of $\delta$. The results are shown in Fig.~\ref{SIfig:numericalsoln} for different rod compressibilities and compared to the incompressible case (dashed line). We observe a linear part with a high slope equal to the compressibility $\beta$ followed by a quasi-plateau very close to the critical buckling load for incompressible rods, independent of $\beta$.
Based on numerical results, we introduce a simplified saturating ramp model. Pure compression before buckling generates a linear relation between loading force, $F$, and displacement, $\delta$: $\sigma = E \epsilon \Rightarrow F = \frac{EA}{H}\delta$. 

\begin{equation}
F(\delta) = EA \left\{f(\delta)-f\left(\delta-\delta_b\right)\right\}
\label{eq:forcemodel1}
\end{equation}

where, $f(u) = \max(0,u) $ is the unit ramp function. The threshold displacement for buckling is given by $\delta_b=\zeta_1^2\frac{EI/H^2}{EA}H=\frac{\zeta_1^2R^2}{4H}$, beyond which the force remains a constant.

We now combine several of these rods to make up the surface force response. Since our sample's flexicles are not uniform in length, they will buckle at different times during indentation. The tallest flexicle will make contact with the indenter first. So, we take that contact point as the origin and the direction of positive displacement be forward indentation. If we denote $\delta^0_i$ as the position of the indenter when it comes into contact with the $i^\text{th}$ flexicle and $H_i$ as the height of the $i^\text{th}$ flexicle, we can modify \eqref{eq:forcemodel1} as:

\begin{equation}
 F_i(\delta) = EA\left\{f(\delta-\delta^0_i)-f\left(\delta-\delta^0_i-\delta^b_i\right)\right\}
\label{eq:forcemodel2}
\end{equation}

where $\delta^b_i=\frac{\zeta_1^2R^2}{4H_i}$. Because of our choice of indenter origin, we can write the $i^\text{th}$ contact position in terms of the $i^\text{th}$ flexicle height as, $\delta^0_i = H_\text{max}-H_i$. If we have $M$ flexicles on our surface, then the collective response of the surface according to the model is simply $F_\text{tot}\left(\delta\right) = \sum_{i}^{M}F_i\left(\delta\right)$.

From experiments, our flexicles are slightly tapered and do not have a uniform cross section. To account for this, we find an equivalent cylinder radius $R = R_{eq}$ for the flexicles by fitting our model to individual flexicle responses to indentation. For the $i^\text{th}$ flexicle, we relate the slope (or the stiffness , $k_i$) of the near-threshold response to the geometric parameters as $k_i = EA_\text{eq}/H_i$, where $A_\text{eq}=\pi R_\text{eq}^2$. We use a Young's modulus of $E = 0.165$ MPa to find $R_\text{eq}$. Using information about the flexicle heights, we use our model to predict a reconstructed force response as shown in~\ref{fig:fig4}.

\begin{figure}
    \centering
    \includegraphics[width=0.5\textwidth]{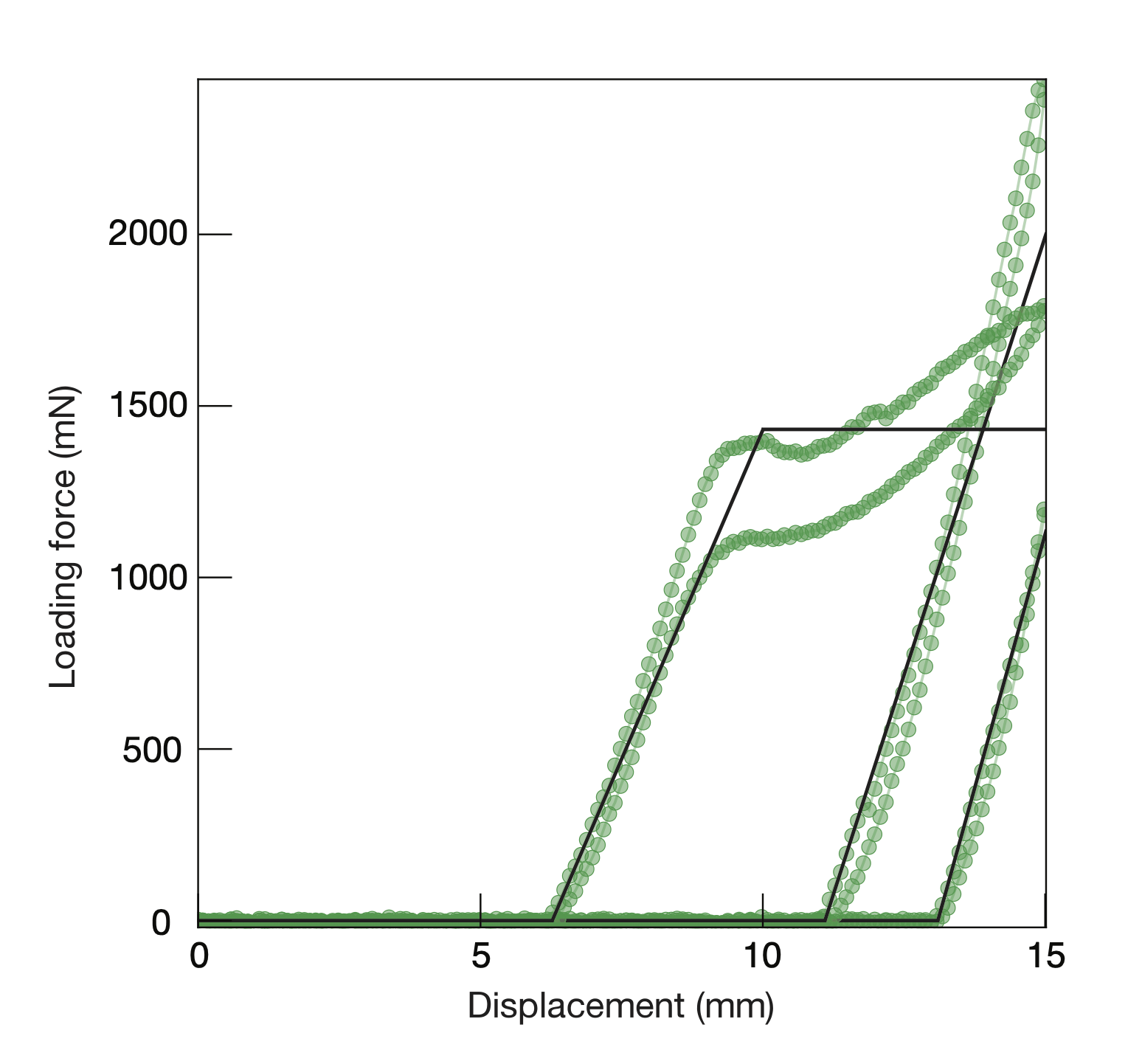}
    \caption{Individual response of 3 different flexicles (green) to indentation, with response prediction by model (black line) overlayed. The model's linear region for the $i^\text{th}$ flexicle has a slope of $E(\pi A_\text{eq})/H_i$, where $H_i$ is the length of the $i^\text{th}$ flexicle.}
    \label{SIfig:SI_compressdata}
\end{figure}

From previous results, we know the statistics behind the heights of our flexicles. Assuming that the flexicle heights belong to the normal distribution, $\mathcal{N}(\overline{H},\sigma_H^2)$ with the probability distribution function $p(H)$, then the number of flexicles with heights between $H$ and $H + \mathrm{d}H$ that the indenter will come into contact with is given by $Mp(H)\mathrm{d}H$ for large $M$. So, for an indentation displacement of $\delta$, we can rewrite the force response in the form of a discrete-sum into an integral:

\begin{equation}
\begin{split}
    \overline{F}_\text{tot}(\delta) = \int_{H_\text{max}-\delta}^{H_\text{max}} &\frac{EA_\text{eq}}{H} \left\{f(\delta-\delta_0(H))\right.\\
    &\left.-f(\delta-\delta_0(H)-\delta_b(H)\right\} Mp(H)\ \mathrm{d}H
\end{split}
    \label{eqn:statforce}
\end{equation}

It follows from the discrete case that the contact point will be $\delta_0(H) = H_\text{max} - H$ and the threshold displacement for buckling will be $\delta_b(H) = \frac{\zeta_1^2R_\text{eq}^2}{4H}$. 

Since we are working with a probability distribution rather than an actual discrete set of flexicle heights, we will take $H_\text{max}$ as the expected value of maximum flexicle height among $M$ flexicles distributed normally as described above. So, $H_\text{max}=\mathbb{E}(H_{\text{max},M}) = \overline{H} + \alpha\sigma_H$, where $\alpha = -\sqrt{2}\left[(1-\gamma_E) \erfc^{-1}\left(2-2/M\right) + \gamma_E \erfc^{-1}\left(2-2/(eM)\right) \right]$ by the Fisher-Tippett theorem~\cite{david2004order} which describes the asymptotic behavior of extreme values in a distribution. $\erfc^{-1}$ is the inverse of the error function, $\gamma_E \approx 0.5772$ is the Euler–Mascheroni constant.

When dealing with acceleration fields higher than gravity, we shrink the length variables in \eqref{eqn:statforce} with the scale factor $\sqrt{a/g}$. By maintaining the area of the flexicle surface as a constant, we accordingly increase the number of flexicles on the surface by a factor of $(a/g)$.


\vspace{4mm}
\section*{Supplementary Information}
\subsection{Spatial arrangement of flexicles}
\label{SI:spatial}
\begin{figure}[!h]
    \centering
    \includegraphics[width=.5\textwidth]{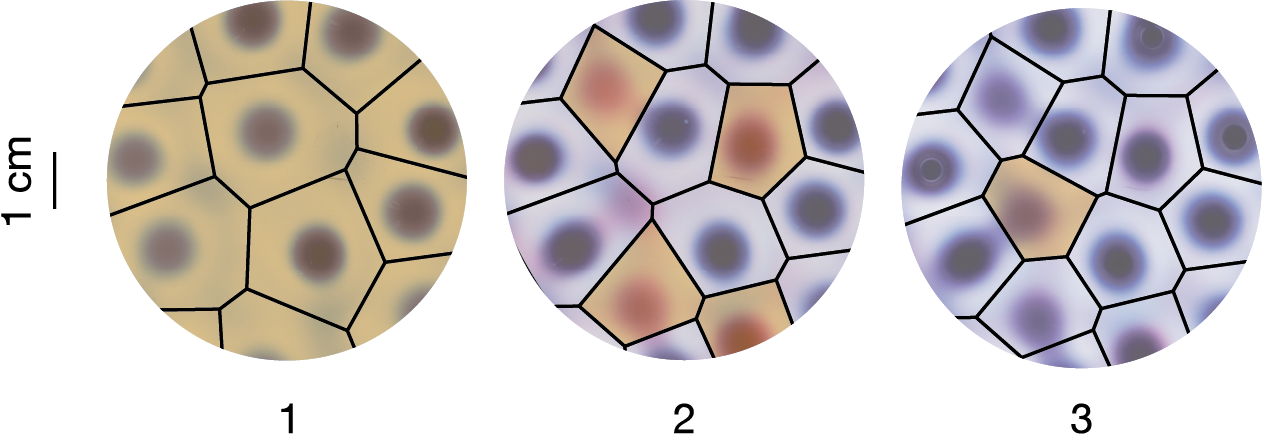}
     \caption{A section of drop lattice after each coating with corresponding Voronoi tesselation overlayed on top. Yellow tiles represent newly formed drops and the other tiles represent previously formed drops.}
    \label{SIfig:fig1}
\end{figure}

Based on the positions of flexicles, we perform a Voronoi tesselation of the surface. Each tile can be characterized using a shape factor, $\xi$ which is defined as $\xi = P^2/4\pi A$, where $P$ and $A$ are respectively, the perimeter and the area of the tile. For an $n$-sided regular polygon, $\xi_n = n/\pi \tan\left(\pi/n\right)$. Using shape factors to describe the tiles allows us to account for their irregularity. In Fig.~\ref{SIfig:fig1}, we show a section of our drop lattices along with their Voronoi tesselations as we add coatings. 

In the first few coatings, while the structures are still relatively less sharp, we note that newer drops are likely to form in the interstitial space between already existing structures. This is indicated by the yellow tiles in Fig.~\ref{SIfig:fig1}. Eventually, as the structures get sharper, we note that it becomes less and less likely for structures to form in the interstitial spaces.

\subsection{Image analysis of the inner structure}
\label{SI:image}
We use \textit{Mathematica}'s Image analysis functions on our cross-section images to extract the individual layers' contours. The images were taken using a photo scanner (Epson Perfection V600 Photo). Interlayer gaps between every layer and its subsequent layer are computed using the obtained layer data. The gap data is smoothed using a moving average over a 200-pixel window size and the points where the layer gap rapidly changes mark the transition between the thin film region and pendant drop region.

\end{document}